\documentclass[12pt]{article}

\usepackage{epsfig}
\usepackage{amssymb}

\newcommand{\app}{\rightarrow}
\newcommand{\half}{\mbox{$\frac{1}{2}$}}

\newcommand{\dt}{\Delta t}
\newcommand{\dx}{\Delta x}
\newcommand{\dzet}{\mathrm{d}\zeta}

\newcommand{\fbf}{{\mathbf{f}}}
\newcommand{\rbf}{{\mathbf{R}}}

\newcommand{\vbf}{{\mathbf{v}}}
\newcommand{\xbf}{{\mathbf{x}}}

\begin{document}

\title{Transition Events in Butane Simulations:\\Similarities Across Models}
\author{Daniel M. Zuckerman$^\ast$ and Thomas B. Woolf$^{\ast \dagger}$\\
$^\ast$Department of Physiology and  $^\dagger$Department of Biophysics,\\
Johns Hopkins University School of Medicine, Baltimore, MD 21205\\
\texttt{dmz@groucho.med.jhmi.edu, woolf@groucho.med.jhmi.edu}
}
\date{DRAFT: \today}
\maketitle

\begin{abstract}
From a variety of long simulations of all-atom butane using both stochastic and fully-solved molecular dynamics, we have uncovered striking generic behavior which also occurs in one-dimensional systems.
We find an apparently universal distribution of transition event durations, as well as a characteristic speed profile along the reaction coordinate.
An approximate analytic distribution of event durations, derived from a one-dimensional model, correctly predicts the asymptotic behavior of the universal distribution for both short and long durations.
\end{abstract}

\newpage
\section{Introduction}
Ensuring appropriate behavior in molecular simulations has long been a topic of interest, and butane has been an important test molecule as the simplest alkane with a dihedral degree of freedom.
In early work on united-atom models Rebertus, Berne and Chandler considered solvent effects on butane conformational equilibrium \cite{Berne-1979}, while Levy, Karplus and McCammon studied stochastic butane dynamics \cite{Levy-1979}.
Pastor and Karplus later investigated the importance of inertial effects in Langevin dynamics, also using a united atom model of butane \cite{Pastor-1989}.
Tobias and Brooks discussed the importance of including hydrogen atoms when considering solvent effects \cite{Brooks-1990}.

One focus of molecular studies has been the dependence of transition rates on the choice of dynamics.
In a study of ethylene glycol, which is polar but also possesses a single dihedral angle, Widmalm and Pastor compared rates and other timescales in molecular and Langevin dynamics \cite{Pastor-1992}.
Similarly, Loncharich, Brooks and Pastor explored the effects of the Langevin friction constant on transitions rates in the larger ``NAMA'' molecule with an eye toward speeding up the generation of conformational ensembles in peptides and proteins \cite{Pastor-1992b}.


Work in one-dimensional systems, in addition to considering the effects of stochastic dynamics on reaction rates (e.g., \cite{Carmeli-1984,Berne-1987b,Hanggi-1990,Voth-1993a,Voth-1993b,Voth-1995}), has explored transition phenomena on still-shorter ``microscopic'' timescales.
Dykman and coworkers have extensively investigated the effects of different noise types and non-equilibrium driving forces on \emph{transition events themselves} in stochastic settings \cite{Dykman-1990,Dykman-1996,Dykman-1997a,Dykman-1997b,Dykman-1998,Dykman-2000}.
The present authors previously considered the influence of the local, non-uniform speed along the reaction coordinate on a ``dynamic importance sampling'' approach to rate computations \cite{Zuckerman-2001}.

Many ``microscopic'' questions have remained unanswered regarding \emph{molecular} transition events.
What is their duration?
Is the reaction coordinate traversed at a uniform speed, or is progress more rapid in some regions?
\begin{figure}[h]

\begin{center}
\epsfig{file=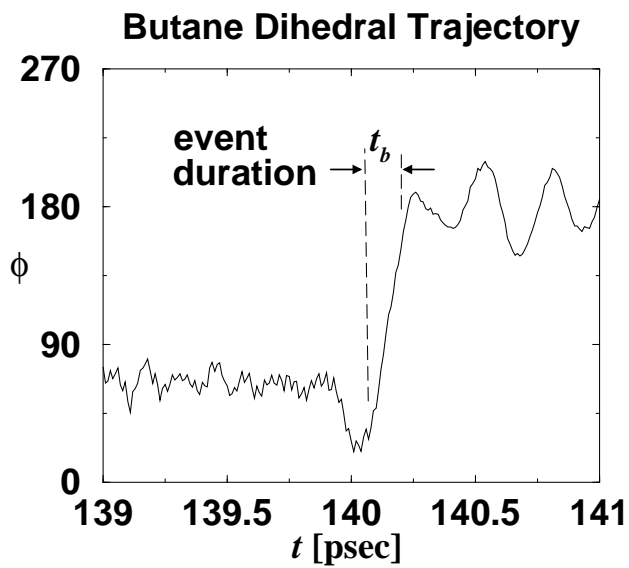,height=3in}
\end{center}

\caption{\label{fig:event}
A single transition event in a butane simulation, illustrating the transition event duration, $t_b$.
The precise definition of $t_b$ is given in Sec.\ \ref{sec:tbresults}
}
\end{figure}
Is there sensitivity to the choice of dynamics or potential?
The need for detailed answers is more pressing in molecular systems because
dynamic importance sampling \cite{Woolf-1998e,Zuckerman-1999,Zuckerman-2001} and ``transition path sampling'' methods \cite{Chandler-1998a,Chandler-1998e} for computing reaction rates require ensembles of transition event trajectories.
A quantitative understanding of transition events could accelerate ensemble generation in these approaches, making larger systems amenable to study.
Moreover, because ensembles of transition events are required for rate estimates, the durations set a fundamental minimum on the computer time required to perform such calculations \cite{Zuckerman-2001}.

The present study provides answers to these questions, using a spectrum of simulations.
First, using the CHARMM potential, all-atom butane is studied in explicit water and acetic acid.
Low- and high-friction Langevin simulations are also performed with CHARMM, along with a separate high-friction run using the AMBER potential.
We compare the butane event-duration distributions and reaction-coordinate speed profiles with those for one dimensional stochastic systems with white and colored noise.

Two interesting generic features emerge.
First, after a simple linear re-scaling by the average, the distribution of transition event durations appears to be \emph{universal} across dynamics, solvent, and potential type --- regardless of whether butane or a one-dimensional system is considered.
The distribution embodies behaviors predicted analytically from a simple one-dimensional model.
Second, the reaction-coordinate speed profiles of butane transitions also display a common, accelerating trend, which matches that of colored-noise, one-dimensional simulations.

In the next section, \ref{sec:models}, we discuss simulation models and methods.
Section \ref{sec:tbtheory} presents an analytic prediction for the distribution of event durations.
Sec.\ \ref{sec:results} contains the simulation results, and we give a summary and some conclusions in Section \ref{sec:conclude}.

\newpage
\section{Simulation Models and Set-Up}
\label{sec:models}
Transition behavior was studied in a variety of settings --- from all-atom, fully solvated molecular dynamics to one-dimensional Langevin dynamics. 
This section describes the protocols and parameters used.

\subsection{Fully-Solvated All-Atom Models of Butane}
To examine solvent effects on barrier crossing behavior, and to ensure a good standard of comparison for stochastic simulations, we performed two explicit solvent simulations of butane.  
In one case the solvent was water and in the other it was acetic acid.  
In both cases the CHARMM program and parameters was used.  
The butane molecule was lightly restrained to the center of the simulation cell using a harmonic, molecule-specific, center-of-mass energy term.
The simulations were performed with constant pressure and temperature and a non-bonded list cut-off of 12 \AA.
The non-bonded list was generated to 15 \AA\ and the van der Waals force switched to zero over a 12 \AA\ distance.  
Electrostatics were shifted over the full 15 \AA\ range to cut-off.  
Dynamics used the leapfrog algorithm (e.g., \cite{Allen-Tildesley}) in CHARMM, which for each particle's position $\xbf$ and velocity $\vbf$ is given by
\begin{eqnarray}
\label{leapfrog}
\xbf(t\!+\!\dt) & = & \xbf(t) + \dt \, \vbf(t \!+ \!\half \dt) \\
\vbf(t \!+ \!\half \dt) & = & \vbf(t \!- \!\half  \dt) + \dt \, \fbf(t) / m
\, ,
\end{eqnarray} 
where $\fbf = -\nabla_{\xbf} U$ for potential energy $U$ and $m$ is the particle's mass.

The acetic acid simulations consisted of 253 explicit molecules of
acetic acid and box dimensions of roughly 65 \AA\ on a side.  
This created
the appropriate density of solvent for a temperature of 300 $K$.  
The water simulations
contained 861 TIP3 water molecules and used a box size of roughly 30 \AA.

\subsection{All-Atom Butane in a Stochastic ``Solvent''}
To maintain as much consistency as possible between the explicit solvent
and the implicit solvent simulations, the stochastic simulations used similar
options to the explicit solvent simulations.  
In particular, the non-bonded energies were cut off at 12 \AA, and a center-of-mass restraint was again used
on the butane molecule.  
Temperature was maintained at 300 $K$.
As would be expected, no periodic images were used and there is no pressure
control in these stochastic simulations.  
Dynamics were performed with the leapfrog Langevin dynamics integrator in CHARMM, namely \cite{Pastor-1992b},
\begin{eqnarray}
\label{langleap}
\xbf(t\!+\!\dt) & = & \xbf(t) + \dt \,\, \vbf(t \!+ \!\half \dt) \\
\vbf(t \!+ \!\half \dt) & = & \vbf(t \!- \!\half  \dt) 
  + \frac{ \xbf(t) - \xbf(t\!-\!\dt)}{\dt} \, \frac{1-\half\gamma\dt}{1+\half\gamma\dt} \nonumber \\
  && + ( \dt / m ) \frac{\fbf(t) + \rbf(t) }{1+\half\gamma\dt} \\
\vbf(t) & = & \sqrt{1+\half\gamma\dt}  
	\, [ \vbf(t \!+ \!\half \dt) + \vbf(t \!- \!\half \dt) ] / 2
\, ,
\end{eqnarray} 
where $\rbf$ is a white-noise frictional force, every component of which is chosen from a Gaussian distribution of zero mean and variance 
$2m\gamma k_B T / \dt $.
We used two values of the friction parameter ($\gamma = 5, 50 \mbox{ ps}^{-1}$) to implicitly model solvent effects;
only the carbon atoms were treated stochastically, while the hydrogen dynamics were governed by the ordinary ($\gamma=0$) leapfrog Verlet algorithm (\ref{leapfrog}).


An additional Langevin simulation was performed using the AMBER potential within the Molecular Modeling Tool Kit (MMTK) \cite{mmtk}.
For this ``overdamped'' (``Brownian'') simulation, dynamics were governed by
\begin{equation}
\label{brown-sim}
\xbf(t+\dt) = \xbf(t) + ( \fbf / m \gamma ) \dt + \Delta \xbf_R \,
\end{equation}
The noise term $\Delta \xbf_R$ was also chosen to be ``white'' --- in this case, selected independently at every time step from a Gaussian of zero mean and variance
\begin{equation}
\label{sigma}
\sigma^2 = 2 \dt k_B T / m \gamma \, .
\end{equation}
All atoms were treated stochastically using the convenient, but we note that overdamped dynamics only make predictions relative to an arbitrary overall timescale \cite{Pastor-1989} embodied here in the (convenient) choice $\gamma = 150$.

\subsection{One-Dimensional Models}
\label{sec:1dmodels}
\begin{figure}[ht]
\begin{center}
\epsfig{file=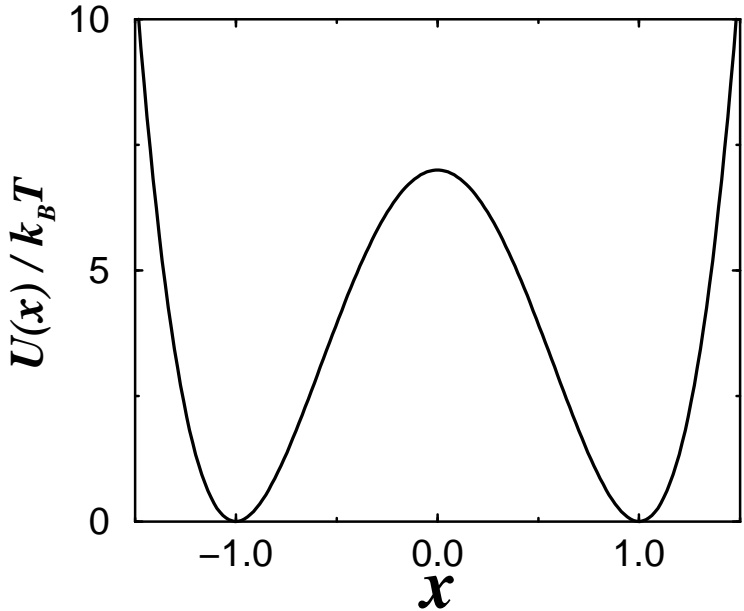,height=3in}
\end{center}
\caption{\label{fig:udw}
The bistable, one-dimensional potential (\ref{udw}) simulated under a variety of conditions discussed in the text.
It is shown here for a barrier height of $E_b = 7 k_B T$ and lengthscale $l\equiv 1$.
}
\end{figure}
We also studied the simple bistable potential discussed in \cite{Zuckerman-2001} and many other places, namely,
\begin{equation}
\label{udw}
U(x) = E_b \, \left[ (x/l)^2 - 1 \right]^2 \; ,
\end{equation}
where $E_b$ is the barrier height and $l$ the length scale.
This potential is shown in Fig.\ \ref{fig:udw}.
All one-dimensional simulations used overdamped Langevin dynamics (\ref{brown-sim}) with ---
for simplicity --- mass, friction, and the thermal energy scale set to unity: $m=\gamma=k_B T\equiv 1$.

Two types of noise were studied.  
The first was uncorrelated ``white'' noise, as discussed above.
We also employed exponentially-correlated noise simulated via an Ornstein-Uhlenbeck process \cite{Gardiner-1985}, in which $\dx_R \equiv y$ is considered a position variable executing \emph{independent} overdamped dynamics (\ref{brown-sim}) in a harmonic potential
$U(y) = (m \gamma / 2 \tau) y^2$ 
subject to white noise with fluctuations given by 
$\sigma_y^2 = (2 \dt / \tau)$.
This formulation yields the auto-correlation function
$\langle y(0) \, y(t) \rangle = \sigma^2 \exp{(-t/\tau)}$,
so that the magnitude of the correlated noise fluctuations is unchanged from the white-noise case.

\newpage
\section{Theoretical Analysis of a Simple Model}
\label{sec:tbtheory}
\begin{figure}[ht]
\begin{center}
\epsfig{file=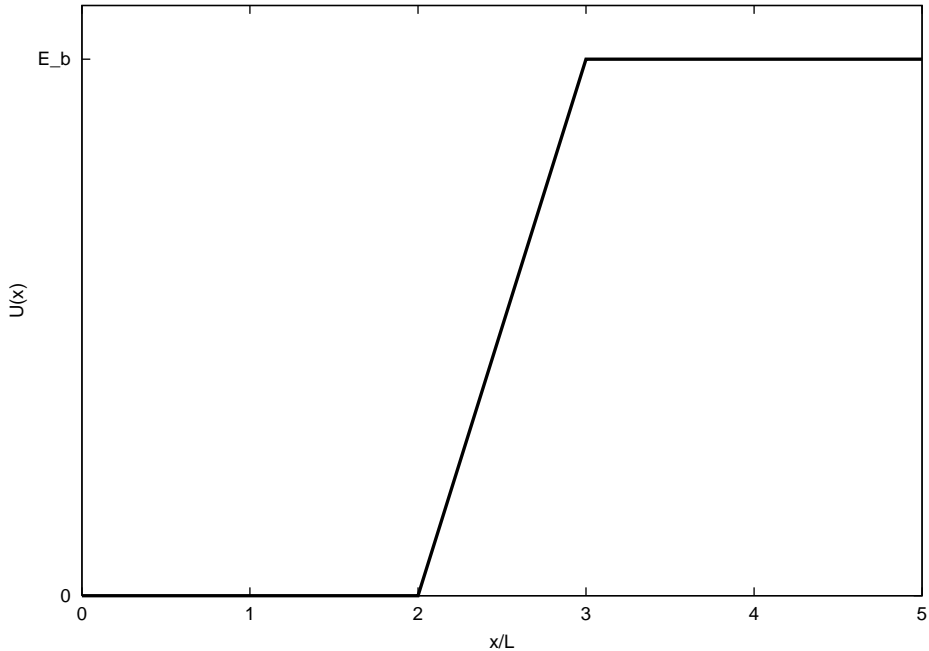}
\end{center}
\caption{\label{fig:incline}
A simple one-dimensional potential used to calculate probabilities of event durations, $t_b$.
}
\end{figure}

This section discusses an extremely idealized potential characterized by a transition-event-duration distribution with \emph{features common among all of the models studied}.
The reader solely interested in the simulation results could safely pass over this discussion and refer to the asymptotic results quoted later.

The one-dimensional potential is shown in Fig.\ \ref{fig:incline}, and the initial state A is defined to be $x < 2$ and state B is $x > 3$.
The motivation for studying such an unphysical model is based on the fundamental observation made in the present report (see next section): because the distribution of durations appears to be universal, it should be derivable from even a trivial model.

The analysis is based on the simulation-step probabilities governing the overdamped dynamics (\ref{brown-sim}) with white noise.
Explicitly, the potential of Fig.\ \ref{fig:incline} is given by
\begin{eqnarray}
\label{uincline}
U(x) & = & 0 \hspace{3cm} ( x < 2 ) \nonumber \\
     & = & (E_b/L) (x-2) \hspace{1cm} ( 2 < x < 3 ) \\
     & = & E_b \hspace{3cm} ( x > 3 ) \; . \nonumber
\end{eqnarray}
For convenience, we introduce the notation 
$\delta x \equiv f \dt / m \gamma = E_b \dt / m \gamma L$ for the inclined region $2 < x/L < 3$.
Because fluctuation increments $\dx_R$ are chosen from a Gaussian distribution, the probability density for choosing an increment $\dx_j = x_{j+1} - x_j$ in the inclined region is given by
\begin{equation}
\label{dxdensity}
T(\dx_j) = \frac{1}{\sqrt{2\pi}\sigma}
	\exp{\left[-(\dx_j - \delta x)^2 / 2 \sigma^2 \right]} \, .
\end{equation}
The probability density of a trajectory of $n_b \equiv t_b/\dt$ steps,
$\zeta^{n_b} = ( x_0, x_1, \ldots, x_{n_b} )$,
is simply the product of single-step densities:
\begin{equation}
\label{pzeta}
Q(\zeta^{n_b}) = \prod_{j=0}^{n_b - 1} T(\dx_j) \, ,
\end{equation}
which, because each of the Gaussian form (\ref{dxdensity}), is properly normalized according to
\begin{equation}
\label{qnorm}
\int \dzet^{n_b} \, Q(\zeta^{n_b}) = 1 \, ,
\end{equation}
where 
$\dzet^{n_b} = \prod_{j=0}^{n_b - 1} \mathrm{d}\dx_j$.

The relative probability of a given event duration, $t_b = n_b \dt$, is a product of two factors:
(i) the total number of probable $n_b$-step trajectories --- whether crossing or not --- multiplied by (ii) the fraction of $n_b$-step trajectories which cross the barrier (i.e., reach the barrier top at $x \geq 3$; see Fig.\ \ref{fig:incline}).
Factor (i) is proportional to $\sigma^{n_b}$, because the fluctuation width in (\ref{dxdensity}) measures the extent of probable steps.
The second factor, the fraction of successful trajectories, may be written formally as
\begin{equation}
\label{crossfrac}
f_b \equiv \int \dzet^{n_b} \, Q(\zeta^{n_b}) \, h_A(x_0) \, h_B(x_{n_b}) \, ,
\end{equation}
where the indicator function $h_Y(x)$ is unity when $x$ is in state Y and zero otherwise.
That is, the probability of a transition event occurring in $n_b = t_b/\dt$ steps is proportional to $\sigma^{n_b} f_b$.

We estimate the fraction of successful trajectories, $f_b$, by considering only those trajectories near to the optimal $n_b$-step crossing trajectory (which differs from the unique, overall-optimal trajectory --- for the the optimal $n_b$ --- discussed in \cite{Zuckerman-2001}).
The optimal $n_b$-step trajectory for the constant-force potential of Fig.\ \ref{fig:incline} consists of uniform steps of length $L/n_b$, and we estimate the number of nearby trajectories as those occurring in a ``fluctuation volume'' $\sigma^{n_b}$ about the optimal;
that is, we estimate
\begin{equation}
\label{crossfracest}
f_b \propto {c_0}^{t_b/\dt} 
\exp{ \left[
-\frac{m \gamma}{2 k_B T} \, t_b \left( \frac{L}{t_b} + \frac{E_b}{m \gamma L}
\right)^2 \right] } \, ,
\end{equation}
where ${c_0}$ is an unknown constant, presumably of order unity.

Finally, the estimate for the probability distribution of event durations is proportional to the product of $\sigma^{n_b}$ and $f_b$, namely,
\begin{equation}
\label{tbdist}
\rho(t_b) \propto  (c_1 \sigma)^{t_b/\dt} 
\exp{ \left[
-\frac{m \gamma}{2 k_B T} \, t_b \left( \frac{L}{t_b} + \frac{E_b}{m \gamma L}
\right)^2 \right] } \, ,
\end{equation}
where $c_1$ is again an unknown constant.
Using the identity $a^n = e^{n \log{a}}$, we can rewrite this approximation as a two-parameter form for purposes of understanding its behavior,
\begin{equation}
\label{tbdistfit}
\rho(t_b) \propto \exp{ \left[
- t_b \left( \frac{c}{t_b} + d
\right)^2 \right] } \, .
\end{equation}
The asymptotic behavior is of particular interest and is given by
\begin{eqnarray}
\label{tbtozero}  \rho(t_b \app 0)      &\sim &\exp{(-c^2/t_b)} \, ,\\
\label{tbtoinfty} \rho(t_b \app \infty) &\sim &\exp{(-d^2 t_b)} \, .
\end{eqnarray}
It is worth noting that all derivatives of $\rho$ vanish as $t_b \app 0$.

\newpage
\section{Results}
\label{sec:results}
\subsection{Transition-Event Durations}
\label{sec:tbresults}
We here present results regarding the durations of transition events, $t_b$;
see Fig.\ \ref{fig:event}.
The universal behavior of the distribution of durations is demonstrated, and limiting behavior is discussed by comparison to the analysis of the previous section.
The average durations are also compared to other timescales which distinguish each system, the correlation time and inverse reaction rate.

\begin{figure}[h]

\begin{center}
\vspace*{0.25cm}
\hspace*{0.1in}\epsfig{file=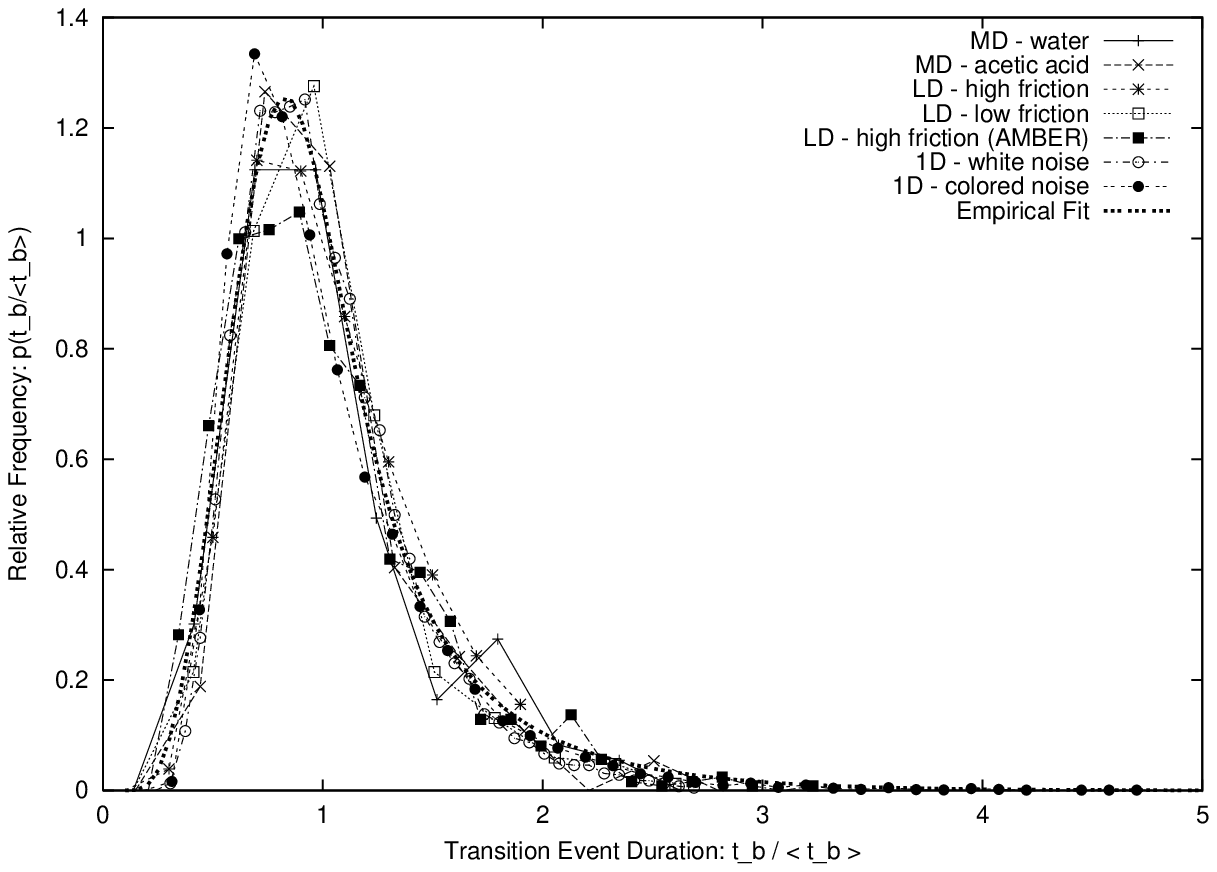,height=2.3in}
\end{center}
\begin{center}
\epsfig{file=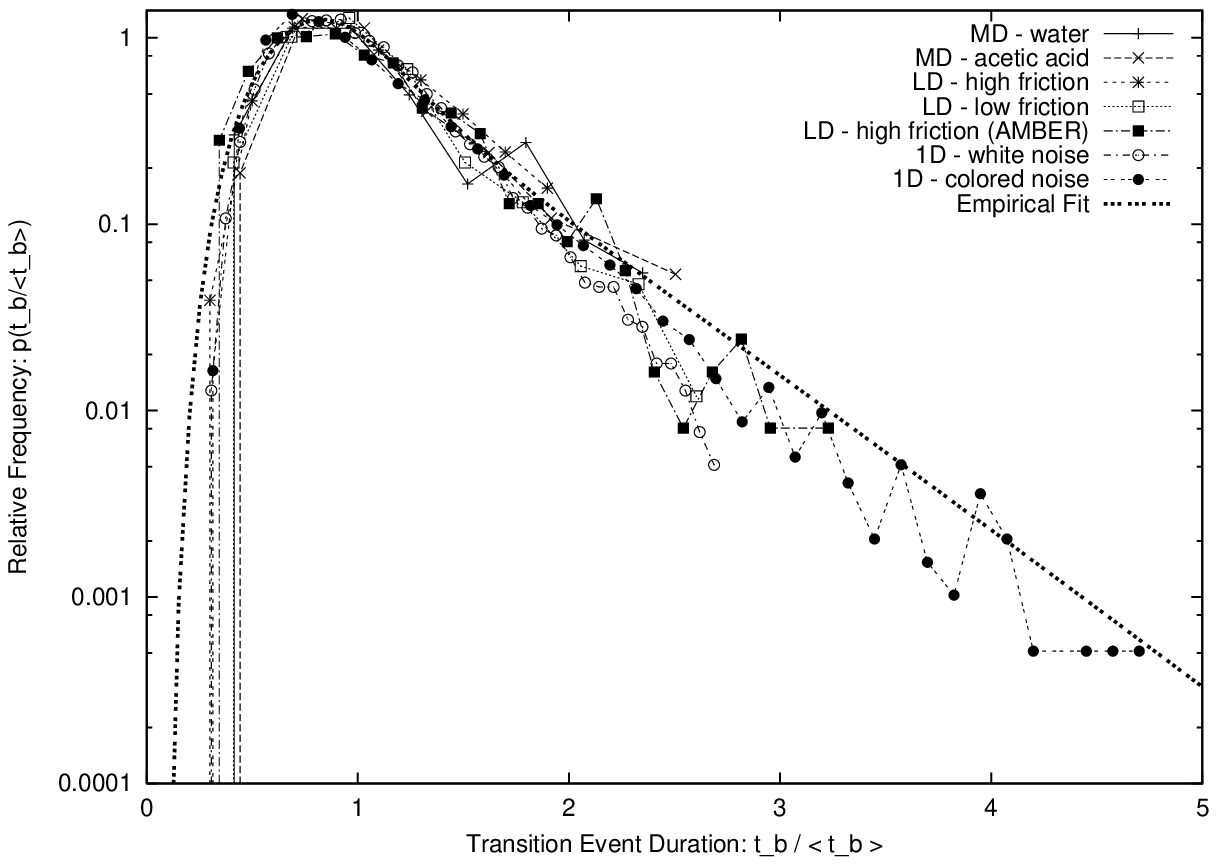,height=2.4in}
\end{center}

\caption{\label{fig:tb-dist}
\small
Collapse of distributions of transition event durations, $t_b$, after simple linear rescaling.
Both butane and one-dimensional simulation data are shown, and
the probability density function, $p$, is plotted for the durations normalized by the average of each data set.
The lower plot, shows identical data on a logarithmic scale to show the common behavior in the \emph{tails} of the distributions.
The butane data sets represent water- and acetic-acid-solvated molecular dynamics (MD) with the CHARMM potential, as well as low- and high-friction Langevin dynamics (LD) with CHARMM, plus a high-friction (overdamped) Langevin simulation using AMBER.
In one dimension (1D), overdamped Langevin dynamics were used, with both white and colored noise.
The empirical fit is given in (\ref{fit}) and discussed in the text.
}
\end{figure}

The definition of a transition duration is necessarily somewhat arbitrary.
Here we used the time between the last exit from the initial state and the first entrance to the final state;
in other words, $t_b$ is the amount of time spent continuously by a trajectory in the transition region.
For transitions between the \emph{gauche}$-$ and \emph{trans} states in butane, we chose the the transition region to be $100 < \phi < 150$;
considering symmetry, the region $210 < \phi < 260$ was chosen for \emph{gauche}$+$ and \emph{trans}.
Following convention, the \emph{trans} state is centered at $\phi=180$.

The principal results for the distributions of event durations are embodied in Fig.\ \ref{fig:tb-dist}.
The simple linear rescalings, by the system-specific average durations $\langle t_b \rangle$, cause the data to collapse onto a fairly well-determined universal curve.
The empirical fit to the data, detailed below, confirms the asymptotic behaviors predicted analytically in Eqs.\ (\ref{tbtozero}) and (\ref{tbtoinfty}):
the gap at small $t_b$ is clearly consistent with $e^{-1/t_b}$ behavior (but not with a more common form like ${t_b}^q e^{-t_b}$ with exponent $q>0$), and the logarithmic scale of Fig.\ \ref{fig:tb-dist}(b) reveals the large $t_b$ behavior to be a simple exponential decay.

The empirical fit shown in Fig.\ \ref{fig:tb-dist} depicts the form
\begin{equation}
\label{fit}
\tilde{\rho}(t_b) = {\cal N}^{-1} \exp{ \left[ -t_b \left(
\frac{ 2 {t_b}^r }{ \tilde{b} + {t_b}^r } + \frac{ \tilde{c} }{{t_b}^2}
\right) \right] } \, ,
\end{equation}
where $\cal N$ is a normalization factor and the three parameters used are
$\tilde{b} = 1.5$, $\tilde{c} = 1.4$, $r = 5$.
The form was designed, by trial and error, both to display the correct asymptotic behaviors and to capture the inflection point to the right of the peak visible on the logarithmic plot, Fig.\ \ref{fig:tb-dist}(b).

\begin{table}[b]
\caption{\label{tab:timescales}
Comparison of timescales for butane dynamics with different explicit and implicit solvents.
The rates, $k$, are between the \emph{trans} ($t$) and \emph{gauche} ($g$) states, and the reciprocals define waiting times between transition events.
For definitions of the dihedral correlation times $\tau_{\mathrm{corr}}$ and $\tau_{\mathrm{decay}}$, see Eq.\ (\ref{taucorr}) and the subsequent text.
The average transition-event duration is denoted by $\langle t_b \rangle$.
All times are in picoseconds.
}
\begin{center}
\begin{tabular}
{l		c	c	l	l	l}
\hline \hline
System 		& $1/k_{g \app t}$ 	
			& $1/k_{t \app g}$ 
				& $\tau_{\mathrm{decay}}$ 
					& $\tau_{\mathrm{corr}}$ 
						& $ \langle t_b \rangle $ \\
\hline
MD/acetic acid	& 55.6	& 294	& 0.40	& 0.02	& 0.136 \\
MD/water	& 35.1	& 115	& 0.21	& 0.04	& 0.145 \\
LD/($\gamma=5$)	& 38.5	& 154	& 0.35	& 0.02	& 0.146 \\
LD/($\gamma=50$)& 32.8	& 151	& 0.06	& 0.076 & 0.200 \\
\hline \hline

\end{tabular}
\end{center}
\end{table}

\subsection{Timescales}
Another question that can be answered based on our data is whether the crossing event durations in a particular system are correlated with other fundamental timescales, like the correlation time and the (inverse) reaction rates.
Table \ref{tab:timescales} apparently answers this question in the negative.
Note that, in the table, the dihedral correlation time is defined by
\begin{equation}
\label{taucorr}
\tau_{\mathrm{corr}} = \langle \, \phi(0) \, \phi(0) \, \rangle^{-1}
  \int_0^\infty \mathrm{d}t \, 
  \langle \, \phi(0) \, \phi(t) \, \rangle,
\end{equation}
where the angular brackets refer here to time averages within the \emph{trans} state only.
Since the dihedral-angle correlation function of the integrand exhibits strong oscillations in molecular dynamics simulations, Table \ref{tab:timescales} also includes the decay time of the exponential ``envelope,'' $\tau_{\mathrm{decay}}$.
The data also suggest that the rates are not simply related to correlation times.

\begin{figure}

\begin{center}
\epsfig{file=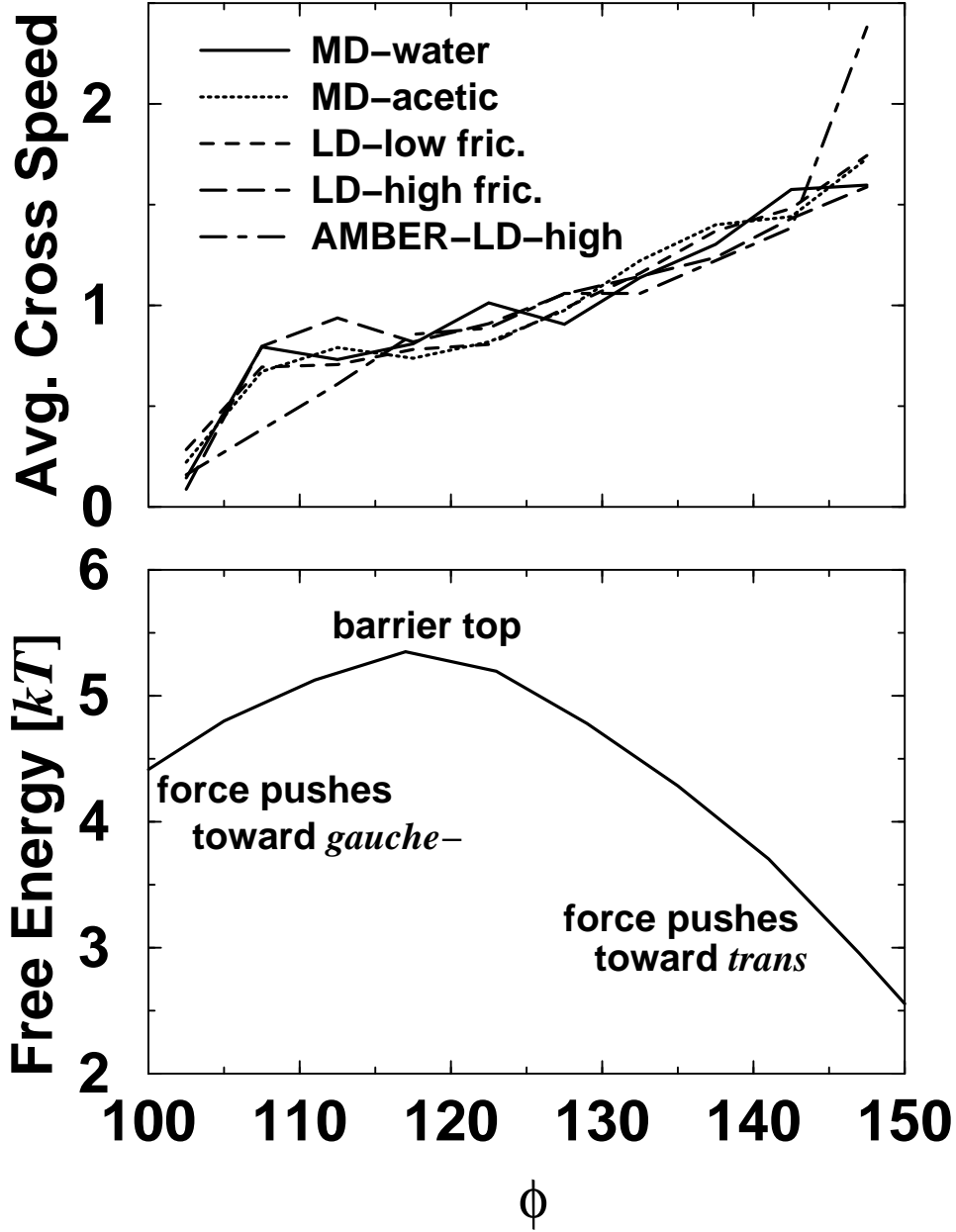}
\end{center}

\caption{\label{fig:butane-cross}
The ``average crossing speed'' during \emph{gauche}$-$ to \emph{trans} transitions in various model butane systems.
We plot the speed ($\Delta \phi / \Delta t$) during transition events \emph{only}, normalized by the \emph{overall} average during these events.
The five butane data sets represent:
(i) water--solvated molecular dynamics with the CHARMM potential;
(ii) acetic-acid--solvated molecular dynamics with CHARMM;
(iii) low-friction Langevin dynamics with CHARMM;
(iv) high-friction Langevin dynamics with CHARMM; and,
(v)  high-friction Langevin dynamics with the AMBER potential.
The absolute crossing speeds can be inferred from the data of Table \ref{tab:timescales}.
For reference, the bottom panel shows the free energy as a function of dihedral angle in the transition region.
}
\end{figure}

\begin{figure}
\begin{center}
\epsfig{file=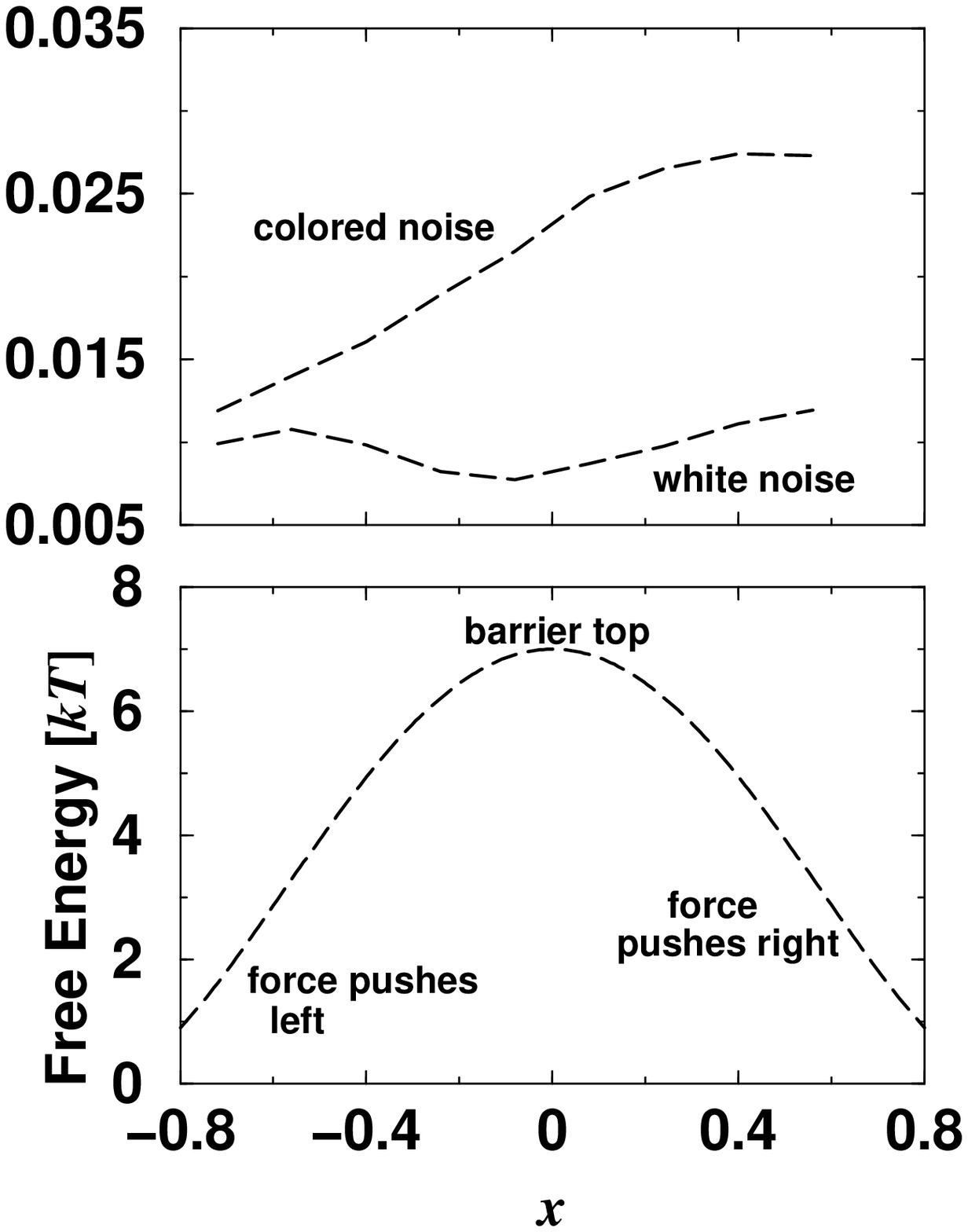}
\end{center}

\caption{\label{fig:1d-cross}
As in Fig.\ \ref{fig:butane-cross}, the reaction coordinate speed profile, now for the one-dimensional bistable well depicted in Fig.\ \ref{fig:udw}.
The average reaction coordinate speeds ($\Delta x/ \Delta t$) are plotted as a function of $x$ for high-friction Langevin simulations, in arbitrary units;
the computations, including colored and white noise, are described in Sec.\ \ref{sec:1dmodels}.  
The ``accelerating'' trend of the colored noise data matches the butane speed profiles, while the symmetric white-noise data do not.
}
\end{figure}

\subsection{Reaction-Coordinate Speed Profiles}
\label{sec:profiles}
The reaction-coordinate speed profile \cite{Zuckerman-2001} depicts the local speed of progress along the reaction coordinate, as exemplified in Fig.\ \ref{fig:butane-cross}.
This sub-picosecond behavior (see Table \ref{tab:timescales}) is uniquely available from simulations, and critically important for dynamic importance sampling methods for rate calculations: see Ref.\ \cite{Zuckerman-2001}.

As in the case of the the crossing time distributions, the results tell a simple and consistent story.
All of the butane simulations show a characteristic ``accelerating'' speed profile during transitions, as indicated by Fig.\ \ref{fig:butane-cross}.
Absolute speeds do differ, as is evident from the average crossing times of Table \ref{tab:timescales}.
The consistent, accelerating profiles from the butane simulations contrast sharply with the white-noise profile of a one-dimensional simulation discussed in detail in Ref.\ \cite{Zuckerman-2001} and also shown in Fig.\ \ref{fig:1d-cross}.

Because even the Langevin butane simulations --- \emph{which were performed with white noise} --- exhibit the accelerating behavior, one can also conclude that molecular fluctuations are inherently colored (i.e., correlated).
The reason is straightforward.
If one focuses on the dynamics of particular butane atom, say a methyl carbon, then according to Eq.\ (\ref{brown-sim}) that atom moves based on the force it feels and a white noise increment.
The force itself, however, reflects the relatively long-timescale motions executed by the whole molecule.
Hence the molecule acts as its own memory, and correlations are not surprising regardless of the type of noise modeled in $\dx_R$.

\newpage
\section{Summary and Conclusions}
\label{sec:conclude}
By comparing butane simulations with explicit and implicit solvents, along with simple white- and colored-noise one-dimensional (1D) systems, we have found a remarkable degree of commonality in short-timescale kinetic properties.
The distributions of transition-event durations $t_b$ \emph{from all systems studied} substantially ``collapse'' onto a single universal function after a simple linear rescaling (Fig.\ \ref{fig:tb-dist}).
Similarly, all the profiles of average local speeds during traversal of the reaction coordinate --- the ``reaction-coordinate speed profiles'' --- of the butane systems behave similarly (Fig.\ \ref{fig:butane-cross}).
These profiles, furthermore, qualitatively match those from 1D colored-noise systems, but are clearly distinct from the 1D white-noise behavior (Fig.\ \ref{fig:1d-cross}).

Despite this universal short-time behavior, quantitative estimates for reaction-rate, correlation, and event-duration timescales (Table \ref{tab:timescales}) reveals no obvious relationship between butane systems with explicit solvents and those with implicit, stochastic ``solvent.''
This suggests that stochastic simulations, at least at the level considered here, cannot be used for quantitative estimates of timescales in fully-solvated computations.
While this assessment may be more pessimistic than that of Widmalm and Pastor \cite{Pastor-1992}, it is not surprising that consideration of an additional timescale, namely $t_b$, makes agreement more difficult.
Perhaps more important are two other issues:
Do simple stochastic representations become better or worse as larger molecules are considered?
What quantity of explicit solvent, say in combination with stochastic boundary conditions \cite{Brunger-1984}, provides a good level of agreement with the array of kinetic quantities in fully-solvated systems?

Further theoretical consideration of the connection between simple and molecular systems would also prove valuable.
In particular, the study of simple models possessing a one-dimensional reaction coordinate coupled nontrivially to simple orthogonal degrees of freedom (based on, e.g., \cite{Carmeli-1984,Berne-1987b,Voth-1993a,Voth-1993b,Voth-1995}) should permit a better understanding of the effects of noise color on the short-time transition behavior considered here.
In a molecular system, as noted in Sec.\ \ref{sec:profiles}, the effective noise on any atom \emph{must} be colored because of the coupling to the non-trivial ``bath'' of other particles.

Following up on the motivation for the investigation reported here, we believe that an appreciation of the results can inform improved sampling approaches for reaction rate computations within the ``transition path sampling'' \cite{Chandler-1998a,Chandler-1998e} and ``dynamic importance sampling'' \cite{Woolf-1998e,Zuckerman-1999,Zuckerman-2001} methods.
These approaches require ensembles of transition-event trajectories.
At a simple level, understanding the transition event duration $t_b$ in molecular systems is critical:
as noted in \cite{Zuckerman-2001}, the duration of molecular crossing events sets, in part, the minimum computation time required to generate a realistic ensemble of transition-event trajectories.
Furthermore, given the universal behavior uncovered here, stochastic simulations evidently reproduce key aspects of explicitly-solvated transition behavior.
Despite the clear importance of solvent effects, this study suggests that trajectories harvested in a computationally inexpensive stochastic context should --- at a minimum --- provide good starting points for refinement in a solvated context.

\section*{Acknowledgments}
A number of people contributed to this project with helpful comments and discussions.
The authors would like to thank Mark Dykman, Lucy Forrest, Hirsh Nanda, Richard Pastor, Horia Petrache, and Jonathan Sachs.
Bernard Brooks provided a useful clarification regarding the Langevin leapfrog algorithm.
Funding for this work was provided by the NIH (Grant GM54782), the Bard Foundation, and the Department of Physiology.
D.M.Z. is the recipient of a National Research Service Award (GM20394).

\newpage

\end{document}